\documentclass[12pt]{article}
\usepackage{epsfig}

\begin{document}

\begin{center}

\vspace*{1.0cm}

{\Large \bf{Heat flow of the Earth and resonant capture of solar 
$^{57}$Fe axions}}

\vskip 1.0cm

{\bf
F.A.~Danevich$^{a}$,
A.V.~Ivanov$^{b}$,
V.V.~Kobychev$^{a}$,
V.I.~Tretyak$^{a}$
}

\vskip 0.3cm

$^{a}${\it Institute for Nuclear Research of the National Academy of 
Sciences of Ukraine, \\ Prospekt Nauki 47, MSP 03680 Kyiv, Ukraine}

$^{b}${\it Institute of the Earth's Crust, Siberian Branch, Russian 
Academy of Sciences, \\ Lermontov St. 128, 664033 Irkutsk, Russia}

\end{center}

\vskip 0.5cm

\begin{abstract}

In a very conservative approach, supposing that \textit{total} heat flow of 
the Earth is exclusively due to resonant capture inside the Earth 
of axions, 
emitted by $^{57}$Fe nuclei on Sun, we obtain limit on mass of hadronic 
axion: $m_a<1.8$ keV. Taking into account release of heat 
from decays of $^{40}$K, $^{232}$Th, $^{238}$U inside the Earth, this 
estimation could be improved to the value: $m_a<1.6$ keV. Both 
the values are less restrictive than limits set in devoted experiments to 
search for $^{57}$Fe axions ($m_a<216-745$ eV), 
but are much better than limits obtained in experiments with $^{83}$Kr 
($m_a<5.5$ keV) and $^{7}$Li ($m_a<13.9-32$ keV). 

\end{abstract}

\section{Introduction}

The general form of the Hamiltonian of quantum chromodynamics (QCD) contains 
a term that violates the CP symmetry in the strong interaction \cite{10, 24}. 
However, this violation is not observed experimentally. For example, only 
upper (and very strict) limit is measured for the neutron electric dipole 
moment, which is related to the CP violating term: 
$d<2.9\times10^{-26}$ $e$$\cdot$cm \cite{1}. This contradiction is known 
as the strong CP problem of QCD. One of the most simple and elegant 
solutions of this contradiction was proposed by Peccei and Quinn in 1977 
\cite{33, 34} by introducing a new global symmetry. The spontaneous violation of 
the PQ symmetry at the energy scale $f_a$ totally suppresses the 
CP violating term in the QCD Hamiltonian. Weinberg \cite{43} and Wilczek \cite{44} 
have independently shown that this model leads to existence of axion -- a 
new pseudo-scalar neutral particle. The mass of axion is related to the 
scale of the PQ symmetry violation: 
$m_a$(eV) $\approx 6\times10^6/f_a$(GeV). The interaction of axion with different 
components of usual matter is characterized by different effective coupling 
constants: 
$g_{a\gamma}$ (interaction with photons), 
$g_{ae}$ (electrons), 
$g_{aN}$ (nucleons), 
which are also 
inversely proportional to $f_a$ and those values are unknown 
(in addition, relations of $g_{a\gamma}$, $g_{ae}$, $g_{aN}$ to $f_a$ 
are model dependent). 

In the first works, the energy of the PQ symmetry violation was considered 
to be close to the scale of the electro-weak symmetry violation and, 
therefore, the axion mass is $\approx$100 keV. But this value of the axion 
mass was soon excluded by experiments with radioactive sources, reactors and 
accelerators (see reviews \cite{1, 3, 10, 23, 24, 29, 37, 38} and references 
therein). Then the standard axion (known as PQWW by names of authors) was 
substituted by other models which allow much bigger values of 
$f_a$ up to the Planck mass of $10^{19}$ GeV: the hadronic axion 
model (KSZV) \cite{22, 39} and the model of the GUT axion (DFSZ) \cite{13, 45}. The 
axion mass and the coupling constants $g_{a\gamma}$, 
$g_{ae}$, $g_{aN}$, which are inversely proportional to 
$f_a$, can have very small values ($m_a$ down to 
$10^{-12}$ eV) in these models, and these axions are sometimes named as 
``invisible''. It should be noted that, besides the solution of the strong 
CP problem, axion is one of the best candidates on the role of the dark 
matter particles. The dark matter, in accordance with the contemporary 
conceptions, constitutes $\approx$23\% of all the matter of the Universe (other 
components are the ordinary baryonic matter $\approx$4\% and the so called dark 
energy $\approx$73\%) \cite{3, 7, 23, 29, 37, 38, 40}.

If axions exist, the Sun can be an intensive source of axions. They can be 
born (1) in the interaction of thermal gamma quanta with fluctuating 
electromagnetic fields within the Sun due to the Primakoff effect and (2) in 
nuclear magnetic transitions in nuclides present in the Sun.

The first effect generates the continuous spectrum of axions with energy up 
to $\sim$20 keV and the mean value of 4.2 keV \cite{41}. The total flux of the 
thermal axions depends on the coupling constant $g_{a\gamma}$ as 
$\phi=(g_{a\gamma} \times 10^{10}$ GeV)$^{2}\times3.5\times 
10^{11}$ cm$^{-2}$ s$^{-1}$. The relation of the axion mass 
$m_a$ to $g_{a\gamma}$ is model dependent; for 
example, this flux is equal (in terms of $m_a$) to 
$\phi=(m_a$/1 eV)$^2\times7.4\times10^{11}$ cm$^{-2}$ s$^{-1}$ 
in the model with GUT axion, whereas other models can possess a 
deeply suppressed axion-photon coupling constant \cite{41}.

In the second effect, de-excitation of excited nuclear levels in magnetic 
(M1) transitions can produce quasi-monoenergetic axions instead of gamma 
quanta, due to axion-nucleon coupling $g_{aN}$. The total energy of 
axions is equal to the energy of gamma quanta. These levels can be excited 
by thermal movement of nuclei (the temperature of the solar core is equal to 
$\sim$1.3 keV, and, therefore, only low-lying levels, like 14.4 keV level of 
$^{57}$Fe or 9.4 keV level of $^{83}$Kr, are excited effectively). Other 
possibility of populating the excited levels is the nuclear reactions in the 
Sun (for example, the 477.6 keV level of $^{7}$Li is populated in the main 
\textit{pp} cycle). 

In spite of theoretical attractiveness of axions, direct experimental 
evidences of their existence are still absent. Indirect astrophysical and 
cosmological arguments give advantage to the axion mass in the range 
$10^{-6}-10^{-2}$ eV or about 10 eV \cite{1, 3, 23, 29, 37, 38}. The 
laboratory searches for axion are based on several possible mechanisms of 
axion interactions with the ordinary matter \cite{3, 23, 29, 37, 38}: 
(1) the inverse Primakoff effect, i.e. conversion of axion to photon in laboratory 
magnetic field (as in the CAST experiment \cite{46}) or in a 
crystal detector (for example, NaI \cite{6}); 
(2) the Compton conversion of axion to photon (analogue of the Compton effect) 
$a + e \to \gamma + e$ \cite{5}; 
(3) the decay of axion to two photons $a \to \gamma \gamma$ \cite{5}; 
(4) the axioelectric effect of interaction with an atom 
$a + (A,Z) \to e + (A,Z)^+$ (analogue of photoeffect) \cite{5}; 
(5) the resonant absorption of axions emitted in nuclear 
M1 transitions in a radioactive source, a nuclear reactor or the Sun by the 
analogue nuclei in a target (see details below). 
It should be noted that 
these mechanisms are based on different kinds of interaction of axion with 
matter, they are sensitive to different coupling constants 
($g_{a\gamma}$, $g_{ae}$, $g_{aN}$), and the 
limits on the values of the constants and on the axion mass are model 
dependent. Thus, diverse experiments are mutually complementary. While the 
most of experiments concern the axion-photon coupling constant 
$g_{a\gamma}$, only the mechanism (5) is related to the 
axion-nucleon constant $g_{aN}$ both in emission and in absorption 
of axion. This allows to exclude uncertainty related to the values of 
$g_{a\gamma}$ and $g_{ae}$.

Below we will discuss the last mechanism in more detail and 
will set the restriction on the axion mass in a conservative assumption that 
all the heat generation within the Earth is caused exclusively by the 
resonant absorption of the solar axions. The possible contribution of 
radioactive decays of $^{40}$K and U/Th families to the heat 
flow will also be taken into account.

\section{The heat generation in the Earth by the resonant capture of solar 
axions}

As mentioned above, axions can be emitted instead of gamma quanta in nuclear 
magnetic transitions (M1) during de-excitation of excited levels of nuclei 
present in the Sun. The corresponding axion flux depends on abundance of 
such nuclei, on radial dependence of the abundance (because of radial 
dependence of temperature which is important for thermal excitation of the 
levels), on the energy of the level (the thermal movement populates 
low-lying levels more effectively), on the mean life of the level and on the 
probability of axion emission instead of gamma quantum in nuclear 
transition. The last value depends on the axion-nucleon coupling constant 
$g_{aN}$ and on the corresponding nuclear matrix elements. The 
calculations should also take into account the probability of absorption of 
the emitted axion in the solar matter.

The expected axion fluxes from thermally excited first levels of $^{23}$Na 
($E_{exc}=440.0$ keV), $^{55}$Mn ($E_{exc}=126.0$ keV), 
$^{57}$Fe ($E_{exc}=14.4$ keV) were calculated in \cite{17}. The axion 
flux from $^{57}$Fe is maximal because of the lowest exciting energy: 
$\phi_{57}=8.5\times10^7\times(m_a/1$~eV)$^2$ cm$^{-2}$ s$^{-1}$ \cite{28}. 
The fluxes from $^{23}$Na and $^{55}$Mn are lower by orders 
of magnitude; they are suppressed by the Boltzmann factor 
$\exp(-E_{exc}/kT)$, where $kT \approx 1.3$ keV in the 
solar center, while the abundances of all 3 nuclides are of the same order. 
The new principle of search for such axions was proposed in \cite{31}: if 
resonant conditions are fulfilled, the solar axion can be captured by a 
respective nucleus (for example, $^{57}$Fe) on the Earth. The particles 
emitted in the subsequent de-excitation of this nucleus (gamma quanta, X 
rays, conversion electrons) can be registered by a suitable detector which 
is placed about the $^{57}$Fe target (or contains $^{57}$Fe nuclei in the 
sensitive volume). The characteristic peak with energy of 14.4 keV would be 
observed in the spectrum of the detector in these conditions.

In the first experiment dedicated to search for the solar $^{57}$Fe axions 
\cite{25}, the iron target (containing 2.1\% of $^{57}$Fe) and Si(Li) detector 
were used. The 14.4 keV peak was not observed, that gave only the upper 
limit on the axion mass: $m_a<745$ eV. This restriction was 
recently improved to values of $m_a<360$ eV \cite{12} and 
$m_a<216$ eV \cite{32}.

Axions supposedly emitted by thermally excited solar $^{83}$Kr nuclei 
($E_{exc}=9.4$ keV) were searched for in the experiment \cite{16}, 
where gaseous proportional counter filled by Kr (11.5\% of $^{83}$Kr) was 
used. The characteristic peak was not observed, and the respective limit on 
the axion mass was $m_a<5.5$ keV.

M1 transitions from the first excited level of $^{7}$Li ($E_{exc}=477.6$ keV) 
can also be a source of quasi-monoenergetic axions \cite{26}. This 
level is populated in the main \textit{pp} chain of nuclear reactions in the 
Sun (which is directly connected to the solar luminosity), when a $^{7}$Be 
nucleus produced by reaction $^3$He + $\alpha$ $\to$ $^7$Be + $\gamma$ 
decays into $^7$Li. In this decay, the 477.6 keV level is populated with 
probability of 10.5\% \cite{15}. In the first experiment \cite{26} on search for such 
axions, a lithium target of 60 g mass and a HP Ge detector were used during 
111 days of measurements; the sought effect was not observed, and only limit 
on the axion mass $m_a<32$ keV was set. The two following 
experiments have improved this limit to $m_a<16$ keV \cite{11} 
and $m_a<13.9$ keV \cite{4}.

The nuclei of $^{7}$Li, $^{23}$Na, $^{55}$Mn, $^{57}$Fe, $^{83}$Kr, etc. can 
be excited by solar axions not only in specially selected targets but 
everywhere. In particular, de-excitation of the resonantly excited levels of 
these nuclei can contribute to the total heat flow from our planet's depths. 
Let us estimate the axion mass in a very conservative assumption that all 
the Earth's thermal flux is caused by resonant absorption of solar axions 
within the Earth.

The directly measured heat flow from both oceanic bottom and 
continents leads to estimation of global power as 
($31\pm1)\times10^{12}$ W \cite{20}. However, it is widely accepted that this value is 
underestimated due to hydrothermal redistribution of heat flow in oceanic 
regions, and model values are used instead: ($44\pm1)\times10^{12}$ W 
\cite{35} or ($46\pm3)\times10^{12}$ W \cite{27}. It is considered that about 
half of this heat is being created by radioactive decays of $^{40}$K and 
nuclei in the chains of $^{238}$U and $^{232}$Th. The distribution of these 
isotopes over the Earth's crust, mantle and core is not exactly known, but 
it is accepted that K, U and Th tend to concentrate in the crust. The 
Earth's internal composition (and the distribution of radioactive nuclei) 
can be investigated by massive (mass of $\sim$1 kt) detectors of antineutrinos 
emitted in nuclear decays within the Earth (the so-called geo-neutrinos). It 
is one of the priority tasks of the modern physics \cite{14}. In particular, the 
Herndon's hypothesis on the nuclear reactor existing in the Earth's center 
\cite{18, 19, 21} can be checked. The heat flow from decays of U/Th/K is 
estimated as $20\times10^{12}$ W (see \cite{14} and references therein); however, 
there exist estimations as high as ($33-43)\times10^{12}$ W \cite{2}.

According to the contemporary conceptions \cite{2}, the Earth consists of the 
crust ($0-35$ km), the upper mantle ($35-660$ km), the lower mantle 
($660-2900$ km), the outer core ($2900-5150$ km), and the inner core 
($5150-6370$ km). The mass of the mantle is near 68\% of the whole Earth's 
mass, it contains 6.26\% of Fe. The core ($\approx$32\% of the Earth's mass) 
includes mainly iron. It is considered that the Earth was formed from the 
primitive matter which had the composition of CI chondrites. Taking this 
model, the authors of \cite{30} calculated that the core contains $78.0-87.5$\% 
of Fe, and all the Earth -- $29.6-32.7$\% of Fe. The chondritic model is 
criticized recently \cite{9, 20, 42}, but the proposed corrections to this model 
do not change significantly the total content of iron in the Earth.

The natural abundance of $^{57}$Fe was measured in samples of different 
origin (crust minerals, magma, different types of meteorites); the 
differences are small \cite{36}, and the recommended value of 
$\delta=2.119$\% \cite{8} can be used for the $^{57}$Fe abundance. Taking the Earth's 
mass of $5.97\times10^{27}$ g \cite{1}, the mass of $^{57}$Fe in the Earth can 
be estimated as ($3.7-4.1)\times10^{25}$ g that corresponds to the 
number of nuclei $N_{57}=(4.0-4.4)\times10^{47}$.

The number of resonant captures of solar axions in a target with 
$N_{57}$ nuclei of $^{57}$Fe per 1 s is equal \cite{12}:

\begin{equation}
R = 4.5 \times 10^{-33} \times N_{57} \times (m_a/1~\mbox{eV})^4.
\end{equation}

The energy of 14.4 keV is released after every capture. This energy is 
totally absorbed in the Earth's body. Taking conservatively the lower 
possible value of the number of $^{57}$Fe nuclei $N_{57}=4.0\times10^{47}$, 
and the highest possible estimation for the Earth's heat flow 
\cite{27} $46\times10^{12}$ W = $2.9\times10^{32}$ eV/s, we equalize the last 
quantity to the power generated by axion captures 
$2.6\times10^{19}\times(m_a/1~\mbox{eV})^4$ eV/s and obtain the upper limit

\begin{equation}
m_a = 1.8~\mbox{keV}.
\end{equation}

The real value of $m_a$ cannot be greater than this value. If one 
takes into account that about half of the heat flow -- $20\times10^{12}$ W 
\cite{14} or even $(33-43)\times10^{12}$ W \cite{2} -- can be generated by 
radioactive decays of $^{40}$K, $^{232}$Th, $^{238}$U within the Earth, the 
estimation (2) can be improved. Subtracting conservatively the lowest 
estimation of radioactive energy generation ($20\times10^{12}$ W) from 
the maximal total Earth's heat flow ($46\times10^{12}$ W) and attributing 
the difference to the heat generation from axion captures, we obtain the 
following upper limit on the axion mass:

\begin{equation}
m_a = 1.6~\mbox{keV}.
\end{equation}

Both the restrictions are few times worse than the limits obtained in 
direct experiments searched for solar $^{57}$Fe axions: $216-745$ eV \cite{12, 25, 
32}. However, they are much better than the limits obtained in the 
experiments with $^{83}$Kr ($m_a < 5.5$ keV \cite{16}) and $^{7}$Li 
($m_a < 13.9-32$ keV \cite{4, 11, 26}). 

\section{Conclusions}

In a very conservative approach, supposing that total heat flow of the 
Earth is exclusively due to resonant capture of axions, emitted by $^{57}$Fe 
nuclei on Sun, we get the limit on the axion mass: $m_a<1.8$ 
keV. Taking into account the heat generated in decays of $^{40}$K, 
$^{232}$Th, $^{238}$U within the Earth, this limit can be improved to 
$m_a<1.6$ keV. Both the values are worse than the limits 
$m_a< 216-745$ eV obtained in direct laboratory searches for 
$^{57}$Fe solar axions \cite{12, 25, 
32} but are much better than the limits obtained in the 
experiments with $^{83}$Kr ($m_a<5.5$ keV \cite{16}) and $^{7}$Li 
($m_a<13.9-32$ keV \cite{4, 11, 26}). Since the rates of both 
emission and resonant capture of axion are governed by the axion-nucleon 
coupling constant $g_{aN}$, the obtained limits do not depend on 
uncertainties in values of the axion-photon ($g_{a\gamma}$) and 
axion-electron ($g_{ae}$) coupling constants.

We used the fact that the flux of monoenergetic solar $^{57}$Fe axions 
should be greater than the fluxes of $^{7}$Li, $^{23}$Na, $^{55}$Mn and 
$^{83}$Kr axions, as well as that the approximately third part of the Earth 
mass is iron. In the following, we plan to improve the obtained limit by 
taking into account the contributions from other possible mechanisms of 
axion interactions with the Earth's matter: the axioelectric effect, the 
Compton axion-photon conversion, the axion decay to two photons, etc.

~

{\bf Acknowledgments.}
The group from the Institute for Nuclear Research (Kyiv, Ukraine)
was supported in part by the Project ``Kosmomikrofizyka''
(Astroparticle Physics) of the National Academy of Sciences of
Ukraine.

\end{document}